\documentstyle[12pt]{article}
\newcommand{\beq}{\begin{equation}}
\newcommand{\eeq}{\end{equation}}

\def\half{{\textstyle{1\over2}}}

\def\p1half{{\textstyle{{{p+1}\over{2}}}}}

\def\23phalf{{\textstyle{{{23-p}\over{2}}}}}

\begin{document}
\thispagestyle{empty}
\begin{titlepage}

\bigskip
\hskip 3.7in{\vbox{\baselineskip12pt
%\hbox{hep-th/0105244}
}}

\bigskip\bigskip\bigskip\bigskip
\centerline{\large\bf Holography and the Canonical Ensemble of
Fermionic Strings}

\bigskip\bigskip
\bigskip\bigskip
\centerline{\bf Shyamoli Chaudhuri
\footnote{shyamoli@thphysed.org}
}
\centerline{214 North Allegheny St.}
\centerline{Bellefonte, PA 16823}
\date{\today}

\bigskip\bigskip
\begin{abstract}
We show that the canonical ensemble in any of the six
supersymmetric string theories, type IIA and IIB, type IB and type
I$^{\prime}$, or heterotic $ E_8$$\times$$ E_8 $ and 
${\rm Spin}(32)/Z_2$ , exhibits a strong version of holography: the
growth of the number of degrees of freedom in the free energy at
high temperatures is identical to that in a two-dimensional
quantum field theory. We clarify the precise nature of the thermal
duality phase transition in each case, confirming that it lies
within the Kosterlitz-Thouless universality class. We
show that, in the presence of Dbranes, and a consequent Yang-Mills gauge
sector, the thermal ensemble of type II strings
is infrared stable, with neither tachyons nor massless
scalar tadpoles. Supersymmetry remains unbroken in the oriented
closed string sector, but is broken by thermal effects in the
full unoriented open and closed type I string theory.
We identify an order parameter for an unusual phase transition in
the worldvolume gauge theory signalled by the short distance
behavior of the pair correlator of timelike Wilson loops. Note Added (Sep
2005).
\end{abstract}

\end{titlepage}

\section{Introduction}

In this paper we will clarify the properties of the thermal
canonical ensemble in the six different supersymmetric string
theories--- type IIA and type IIB, heterotic $E_8$$\times$$E_8$
and ${\rm Spin}(32)/{\rm Z}_2$, type I and type I$^{\prime}$. The
Polyakov path integral over worldsheets offers a first principles
approach to the string canonical ensemble, originally pointed out
in \cite{poltorus}. It is important to appreciate the significance
of this observation. Recall that we only know how to formulate
perturbative string theory in a given spacetime background: a
\lq\lq heat bath", represented by the given spacetime metric and
background fields, is forced upon us, {\em implying that a
self-consistent treatment of perturbative string thermodynamics
must necessarily be restricted to the canonical ensemble of
statistical mechanics}. Thus, while an immense, and largely
conjectural, literature exists on proposals for microcanonical
ensembles of weakly-coupled strings \cite{ea,micro}, the
conceptual basis for such discussion is full of holes. Some of
these pitfalls were already pointed out in \cite{aw}. Rather than
engage in such conjectural debate, we will instead postpone
treatment of the microcanonical ensemble until when we have
consensus on a nonperturbative, and background independent,
formalism for string/M theory. Matrix proposals for M theory may
eventually offer such a possibility. It should be kept in mind
that, strictly speaking, the {\em microcanonical} ensemble is what
is called for when discussing the thermodynamics of the Universe
\cite{hawk}: the Universe is, by definition, an isolated closed
system, and it is meaningless to invoke the canonical ensemble of
the \lq\lq fundamental degrees of freedom". But there are many
simpler questions in braneworld cosmology that might indeed be
approachable within the framework of string thermodynamics in the
canonical ensemble.

\vskip 0.1in In what follows, we will both review, and make
important corrections to, the standard worldsheet derivations for
the canonical ensemble of the different supersymmetric string
theories at finite temperature, following the approach outlined in
Polchinski's 1986 treatment of the canonical ensemble of closed
bosonic strings.\footnote{In the earlier papers
\cite{bosonic,decon}, we have pointed out a peculiar ambiguity in
the Euclidean time prescription for finite temperature theories
illustrated by its application to the canonical ensemble of free
strings. To avoid the misconception that any of the results in our
current paper are tied to the question of whether Euclidean time
has the topology of a circle or that of an interval, we will
refrain from any mention of the ambiguity in this paper. {\em We
emphasize that all of the derivations in the current paper make
the standard assumption that Euclidean time has the topology of a
circle, with inverse temperature identified as follows:
$\beta$$=$$2\pi r_{\rm circ.}$.}}Given the enormous literature to
date on string thermodynamics, even within the framework of the
canonical ensemble, it may be helpful to clarify the precise
corrections made in our analysis to the previous standard
treatments. Polchinski gave the original derivation of the free
energy of the canonical ensemble of free closed bosonic strings
using the string path integral in \cite{poltorus}. However, in
physically interpreting the correctly derived expression, he omits
to mention that there are low temperature tachyonic thermal
momentum modes in the spectrum {\em at all temperatures starting
from zero} \cite{bosonic}. Thus, Polchinski's oft-quoted
identification of the first winding mode instability with the
onset of a \lq\lq Hagedorn" phase transition is suspect: the free
energy is already ridden with tachyonic divergences long before
the \lq\lq Hagedorn" temperature has been reached \cite{bosonic}.
This misleading identification has also been made independently in
the papers by Kogan, and by Sathiapalan, listed in \cite{tan}.

\vskip 0.1in The canonical ensembles of the type II and heterotic
strings were discussed in detail by Atick and Witten in \cite{aw}.
They were the first authors to introduce (??) thermal mode number
dependent?? temperature-dependent phases in the expression for the
free energy, except that the particular choice type II
heterotic... given in their paper {\em violates modular
invariance}. As emphasized in the original derivation closed bos
in \cite{poltorus}, and in our reviews in \cite{bosonic,decon},
the string path integral provides a first-principles derivation of
the free energy of the canonical ensemble of strings, as
calculated in string loop perturbation theory: the free energy is
precisely given by the generating functional of connected vacuum
graphs. Notice that the one-loop analysis is not subject to the
Jeans instability \cite{aw}: the strings do not gravitate at this
order in perturbation theory. Thus, a self-consistent description
of the canonical ensemble of free strings at the one loop level is
fully compatible with the worldsheet formalism and had already, in
fact, been given by Polchinski in \cite{poltorus}. As befits any
perturbative closed string amplitude, {\em the expression for the
free energy is therefore required to be modular invariant from
general considerations of two-dimensional reparameterization and
Weyl invariance in perturbative string theory}.

\vskip 0.1in We should also note that \cite{aw} makes an incorrect
assertion about thermal duality: it is indeed true that the free
energy of any sensible physical system should be a monotonically
increasing function of temperature, but there is nothing \lq\lq
unphysical" about the generating functional of connected vacuum
graphs being thermal self-dual. The Polyakov path integral yields
$W(\beta)$, {\em not} $F(\beta)$, and $W(\beta)$ is indeed thermal
self-dual in the case of the closed bosonic string theory: recall
that a thermal duality transformation is simply a Euclidean
timelike T-duality transformation. The free energy of the string
ensemble is given by $F(\beta)$$=$$-W(\beta)/\beta$, which is,
happily, a monotonically increasing function of temperature.
Thermal duality has been extensively explored in the early days of
string theory, notably by E.\ Alvarez and collaborators
\cite{ea}.\footnote{These authors also attempt to \lq\lq derive"
the properties of the microcanonical ensemble from those of the
canonical ensemble \cite{ea,micro} applying them to models for
superstring cosmology. We have already noted that such attempts
are conceptually flawed. The technical reason why the Legendre
transform linking the microcanonical and canonical closed string
ensembles cannot be carried out in closed form is that the
integration over the moduli of the Riemann surfaces summed in the
expression for the free energy cannot be carried out explicitly.}
Duality also makes an appearance in several of the papers listed
in \cite{tan}. But the fundamental significance of a duality phase
transition in the canonical string ensemble characterized by the
analyticity in temperature of an {\em infinite} hierarchy of
thermodynamic potentials appears not to have been noticed prior to
our work \cite{bosonic}. It is this analytic behavior which
provides the unambiguous signature of a phase transition belonging
to the Kosterlitz-Thouless universality class \cite{kt},
exemplified by our analysis in section 3 of the heterotic string
ensemble. Finally, it should be emphasized that it is only the
vacuum functional of string theory that is thermal self-dual in
certain cases, namely, closed bosonic string theory. In the cases
described in this paper, thermal duality acts within a pair of
supersymmetric string theories, mapping the vacuum functional of
one to the other. But in all cases, the free energy and remaining
thermodynamic potentials, given by the basic thermodynamic
identities listed in Eq.\ (\ref{eq:freene}), are most certainly
{\em not} thermal self-dual.

\vskip 0.1in Our results for the canonical ensemble of the type I,
or type I$^{\prime}$, unoriented open and closed string theory
with Dbranes are entirely new. Also new is the observed connection
between the spontaneously broken thermal duality, and the signal
for a phase transition to a {\em holographic} phase at high
temperatures: the growth with temperature in the free energy of
the type I$^{\prime}$ ensemble at high temperatures is that of a
{\em two-dimensional} field theory. Our results bring to
completion an unambiguous demonstration of the holographic
behavior of {\em all} of the perturbative string theories: type
I-I$^{\prime}$, heterotic, and type IIA-IIB, in addition to that
of the closed bosonic string theory first demonstrated in
\cite{polchinskibook}, following the conjecture in \cite{aw}. The
notion of a holographic principle has played an active role in
recent developments in String/M theory \cite{sussk,bousso}. But it
should be emphasized that the holographic growth of the free
energy at high temperature in string theory is a much more drastic
reduction in the number of degrees of freedom: $F$ $\propto$ $T^2$
for the ten-dimensional string theories, not simply as fast as an
area, $\propto$ $T^9$, rather than as a volume, $\propto$
$T^{10}$, as postulated in \cite{sussk,bousso}. Type I string
thermodynamics was not discussed in \cite{poltorus,aw}, although
discussions have appeared elsewhere in the literature. To the best
of our knowledge, all of these references neglect the errors we
have pointed out in the standard references \cite{poltorus,aw}.
Treatments which are restricted to the supergravity field theory
limit--- treating the Dbranes as semi-classical sources in a
thermal bath of gravitons, dilatons, and nonabelian gauge fields,
should be viewed as effective descriptions of low energy thermal
string/M theory at finite temperature.

\vskip 0.05in All of the analysis in this paper, and in
\cite{bosonic,decon}, is based on the one-loop amplitude in string
theory, a loop contribution to the thermodynamic potentials that
is independent of the strength of the string coupling constant
\cite{zeta}. The reader may wonder why we omit mention of {\em
tree}-level contributions to the vacuum energy density. In the
case of pure closed string theories, or for the closed string
sector of the type I and type I$^{\prime}$ open and closed string
theories, there are none \cite{polchinskibook}: the underlying
reason can be traced to reparametrization invariance of
worldsheets with the topology of a sphere. On the other hand, in
the presence of Dbranes, one indeed finds a nontrivial tree-level
contribution to the vacuum energy density from the disk amplitude,
namely, the Dbrane tension \cite{dbrane,polchinskibook,zeta}. The
reader can find a nice discussion of the tree-level contributions
to the vacuum energy density in these references.

\vskip 0.1in The thermodynamics of an ensemble of strings is not
the same as that of an ensemble of infinitely many free particle
species with Planck scale masses. While there is a
well-established back-of-the-envelope argument deriving the
existence of a Hagedorn divergence in {\em particle} ensembles at
finite temperature described by a Boltzmann distribution
\cite{hagedorn} (see the review in the appendix of
\cite{bosonic}), we must be careful not to apply this to the
string ensemble. The partition function, ${\rm Z}(\tau_i)$, on a
strip of given shape and size, namely, with fixed values of the
worldsheet moduli, $\tau_i$, indeed describes a conformally
invariant two-dimensional field theory with an exponentially
rising density of states. But the vacuum functional in string
theory is given by a reparametrization invariant {\em integral}
over strips of all possible shape and size. Remarkably, the
physical state spectrum of string theory turns out to contain many
fewer states than in the two-dimensional field theories described
by the {\em integrand} of the one-loop vacuum amplitude as a
consequence of reparametrization invariance
\cite{bosonic,zeta,relevant}.

\vskip 0.1in In this paper, we will perform a first-principles
analysis of the generating functional of connected one-loop vacuum
string graphs, $W(\beta)$ $\equiv$ ${\rm ln}$ ${\rm Z}(\beta)$,
following the approach given in \cite{polyakov,poltorus,zeta}. The
vacuum energy density can, of course, be directly inferred from
$W(\beta)$. Let us recall the basic thermodynamic identities of
the canonical ensemble \cite{pippard}:
\begin{equation}
F= -W/\beta = V \rho , \quad P =
      - \left ( {{\partial F}\over{\partial V}} \right )_T , \quad
      U = T^2 \left ( {{\partial W}\over{\partial T}} \right )_V
, \quad
      S = - \left ( {{\partial F}\over{\partial T}} \right )_V ,
 \quad
      C_V = T \left ( {{\partial S}\over{\partial T}} \right )_V
\quad .
\label{eq:freene}
\end{equation}
Note that $W(\beta)$ is an intensive thermodynamic variable
without explicit dependence on the spatial volume. $F$ is the
Helmholtz free energy of the ensemble of free strings, $U$ is the
internal energy, and $\rho$ is the finite temperature effective
potential, or vacuum energy density at finite temperature. $S$ and
$C_V$ are, respectively, the entropy and specific heat of the
thermal ensemble. The pressure of the string ensemble simply
equals the negative of the vacuum energy density, as is true for a
cosmological constant, just as in an ideal fluid with negative
pressure \cite{peebles,pippard}. The enthalpy, $H$$=$$U$$+$$PV$,
the Helmholtz free energy, $F$$=$$U$$-TS$, and the Gibbs function,
also known as the Gibbs free energy, is $G$$=$$U$$-$$TS$$+$$PV$.
As a result of these relations, {\em all} of the thermodynamic
potentials of the string ensemble have been give a simple,
first-principles, formulation in terms of the path integral over
worldsheets. Notice, in particular, that since $P$$=$$-\rho$, the
one-loop contribution to the Gibbs free energy of the string
ensemble vanishes identically!

\vskip 0.1in Let us summarize our results. While the zero
temperature type IIA or type IIB string is supersymmetric and
tachyon-free, we will show that the thermal spectrum of either
type II string theory contains an infinite tower of tachyonic
physical states at infinitesimal temperature: this pathological
behavior is already indicated by the preliminary analysis in
\cite{aw}, raising the puzzling question of whether one can
describe a stable thermal ensemble of type II strings in flat
noncompact 10d spacetime? It turns out, as unambiguously clarified
by us in an accompanying work using the worldsheet renormalization
group (RG) and the g-theorem \cite{relevant,al}, that the type IIA
vacuum at temperatures below $T_{w=1}$, namely, the temperature at
which the first of the winding modes turns tachyonic, is {\em
unstable}, and the flow of the worldsheet renormalization group is
{\em towards} the noncompact and supersymmetric zero temperature
vacuum. How, then, does one describe a stable thermal ensemble of
type IIA strings? The key to this puzzle lies in the Ramond-Ramond
sector which has, thus far, not been taken into account. To
understand why a nontrivial Ramond-Ramond sector might enable the
elimination of tachyons from the thermal spectrum at all
temperatures, let us turn to the heterotic and type I string
ensembles.

\vskip 0.1in As we show in section 2 and 3, the key feature
necessary for a tachyon-free thermal ensemble that is also
self-consistent with the thermal duality relations of String/M
theory, lies in introduction of a temperature dependent Wilson
line: an option available in any string theory with Yang-Mills
gauge fields. As a consequence, it will turn out that both the
heterotic $E_8$$\times$$E_8$ and ${\rm Spin}(32)/{\rm Z}_2$
theories, and the type I and type I$^{\prime}$ $O(32)$ theories,
have a {\em tachyon--free finite temperature ground state at all
temperatures starting from zero, and with gauge group
$SO(16)$$\times$$SO(16)$.} From the perspective of the low energy
finite temperature gauge-gravity theory \cite{bernard}, the
presence of the temperature dependent Wilson line can be
interpreted as quantization in the modified axial gauge, $A_0$ $=$
constant, where the constant has been chosen to be {\em
temperature dependent}. In fact, it is straightforward to write
down the correct result for the vacuum functional at finite
temperature if we recall some old results in the heterotic string
literature. In \cite{sw,agmv,dh,klt}, the possible choices of spin
structure for the worldsheet fermions consistent with both the
$(1,0)$ superconformal invariance and modular invariance of the
heterotic closed string theory were studied, providing a
classification of ten-dimensional heterotic strings. There is a
{\em unique} solution which is both nonsupersymmetric and
tachyon-free, and it has nonabelian gauge group
$O(16)$$\times$$O(16)$. In an interesting follow-up work
\cite{ginine}, Ginsparg studied the compactification of the
supersymmetric $E_8$$\times$$E_8$ string on a circle,
demonstrating, thereby, that by tuning both the radius and a
possible Wilson line, it was possible to interpolate smoothly
between the circle-compactified $E_8$$\times$$E_8$ and
$Spin(32)/{\rm Z}_2$ string theories. Related developments in this
period were the interesting numerical studies \cite{gv,itoyama} of
the effective potential in nonsupersymmetric ground states of the
heterotic string, examining the effects of tuning the spatial
radi, $r_i$, background antisymmetric field, $b_{ij}$, or Wilson
lines, $A_i$ \cite{nsw}. It should be mentioned that the
nonsupersymmetric and tachyon-free $O(16)$$\times$$O(16)$ string
also makes an appearance in recent work on the interpolation
between the circle-compactified $E_8$$\times$$E_8$ and
$Spin(32)/{\rm Z}_2$ theories, in the context of a conjectured
strong-weak heterotic-type I duality in the absence of
supersymmetry \cite{dienes}. We will be able to make contact with
some of these results in section 3, especially the analysis of
Wilson lines that appears in the interpolation studied in
\cite{ginine}.

\vskip 0.1in The plan of this paper is as follows. Section 2
begins with a derivation of the free energy for the canonical
ensemble of heterotic strings. We show that in the presence of a
temperature-dependent Wilson line background, the thermal spectrum
is tachyon-free at all temperatures starting from zero and has
gauge symmetry $SO(16)$$\times$$SO(16)$. In section 2.2, we derive
expressions for the hierarchy of thermodynamic potentials of the
canonical ensemble, demonstrating both a holographic duality
relation, and the signature of a Kosterlitz-Thouless duality
transition \cite{kt}. We clarify that both the free energy and
entropy are finite with no divergences. {\em In particular, there
is no evidence for a Hagedorn phase transition in the tachyon-free
heterotic string ensemble}.

\vskip 0.1in In section 3 we introduce the type I and type II
unoriented open and closed string theories with Dbranes. We begin
with the closed oriented sector of this theory, computing the
one-loop vacuum functional for an equilibrium ensemble of either
type IIA or type IIB closed oriented strings at finite
temperature, with a trivial Ramond-Ramond sector. In Section 3.2,
we address the low temperature thermal instability of either type
II ensemble in the absence of Dbranes. The instability of such an
ensemble, and the worldsheet renormalization group flow to the
supersymmetric infrared fixed point at zero temperature, is
described in \cite{relevant}. Thus, the closed oriented sector of
the type I or type II theories does not break supersymmetry as a
consequence of thermal effects. Section 3.3 describes the
computation of the one-loop free energy for the remaining sectors
of the type IB unoriented open and closed string theory. We derive
the generating functional of one-loop vacuum string graphs, having
noting that the contribution from worldsurfaces of torus topology
vanishes. Remarkably, we find that the one-loop contribution to
the free energy from the {\em sum} of the annulus, Mobius Strip,
and Klein Bottle topologies also vanishes as a consequence of
requiring tadpole cancellation for the unphysical Ramond-Ramond
scalar, despite the fact that supersymmetry has been broken! It is
is most intriguing to find a nonsupersymmetric type I ground
state, with supersymmetry broken by finite temperature effects,
despite a vanishing one-loop vacuum energy density.

\vskip 0.1in The holographic growth of the number of degrees of
freedom in type I string theory at high temperatures is
demonstrated in section 4.1, where we also evaluate the scaling
behavior for the first few thermodynamic potentials, verifying
that the duality phase transition is in the Kosterlitz-Thouless
universality class \cite{kt}. Finally, in section 4.2, we find a
plausible order parameter for a phase transition in the
worldvolume gauge theory by looking for a signal in the {\em short
distance} behavior of the pair correlator of closed timelike
Wilson loops. We find a transition temperature at $T_C$, or at a
temperature slightly below $T_C$ in the presence of an external
electromagnetic field. We pause to remark that this continuous
phase transition is the {\em only} possible phase transition
compatible with the {\em analyticity of string theory amplitudes}
and consequently, accessible within the worldsheet formalism of
perturbative string theory. The intuition that a gas of strings
can transition into a high temperature long string phase is an old
piece of string folklore \cite{polchinskibook,sussk,garyp},
although usually formulated within the context of a model for the
microcanonical ensemble. It is possible that our results can be
interpreted as evidence for such a phase transition. Conclusions
and a brief discussion of open questions appear in section 5.

\section{Heterotic Closed String Ensemble}

\vskip 0.1in The heterotic closed string theory is possibly the
closest supersymmetric string analog of the closed bosonic string
theory considered by Polchinski in \cite{poltorus}, so let us
begin with this case. Consider the ten-dimensional supersymmetric
$E_8$$\times$$E_8$ theory at zero temperature. The
$\alpha'$$\to$$0$ low energy field theory limit is 10D $N$$=$$1$
supergravity coupled to $E_8$$\times$$E_8$ Yang-Mills gauge
fields. What happens to the supersymmetric ground state of this
theory at finite temperatures? Assuming that a stable thermal
ensemble exists, the finite temperature heterotic ground state
with nine noncompact spatial dimensions is expected to be
tachyon-free, while breaking supersymmetry. Moreover, consistency
with the low energy limit, which is a finite temperature
gauge-gravity theory, implies that the thermal string spectrum
must contain Matsubara-like thermal momentum modes. But the
thermal spectrum is also likely to contain winding modes as
expected in a closed string theory \cite{poltorus}. Most
importantly, since we are looking for a self-consistent string
ground state with good infrared and ultraviolet behaviour, it is
important that the one-loop vacuum functional preserve the usual
worldsheet symmetries of $(1,0)$ superconformal invariance and
one-loop modular invariance. Finally, all of our considerations
are required to be self-consistent with thermal duality
transformation defined as a Euclidean timelike T-Duality
transformation. Since the action of spatial target space dualities
on the different supersymmetric string theories are extremely
well-established, the finite temperature vacuum functional is
required to interpolate between the following two spacetime
supersymmetric limits: in the $\beta$$\to$$\infty$ limit we
recover the vacuum functional of the supersymmetric
$E_8$$\times$$E_8$ heterotic string, while in the $\beta$$=$$0$
limit we must recover, instead, the vacuum functional of the
supersymmetric $\rm Spin(32)/Z_2$ heterotic string. The reason is
that the $E_8$$\times$$E_8$ and $\rm Spin(32)/Z_2$ heterotic
string theories are related by the Euclidean timelike T-duality
transformation: $\beta_{E_8 \times E_8}$$\to$$\beta_{\rm
Spin(32)/Z_2}$$=$$4\pi^2 \alpha'/\beta_{E_8 \times E_8}$.

\subsection{Axial Gauge and the Euclidean Timelike Wilson Line}

\vskip 0.1in  The generating functional of connected one-loop
vacuum string graphs is given by the expression:
\begin{equation}
W_{\rm het.} = L^9 (4\pi^2 \alpha^{\prime})^{-9/2} \int_{\cal F}
{{d^2 \tau}\over{4\tau_2^2}}
  (2 \pi \tau_2 )^{-9/2} |\eta(\tau)|^{-14} Z_{\rm het.} (\beta)
\quad . \label{eq:het}
\end{equation}
The canonical ensemble of oriented closed strings occupies the
box-regularized spatial volume $L^9$. The thermodynamic limit is
approached as follows: we take the limit
$\alpha^{\prime}$$\to$$0$, $L$$\to$$\infty$, holding the
dimensionless combination, $L^9 (4\pi^2 \alpha^{\prime})^{-9/2}$,
fixed. The function $Z_{\rm het.}(\beta)$ contains the
contributions from the rank (17,1) Lorentzian self-dual lattice
characterizing this particular ground state of the circle
compactified $E_8$$\times$$E_8$ heterotic string. Thus, we wish to
identify a suitable interpolating expression for $W_{\rm het}$
valid at generic values of $\beta$, matching smoothly with the
known vacuum functional of the supersymmetric $E_8$$\times$$E_8$
string theory at zero temperature ($\beta$$=$$\infty$):
\begin{eqnarray}
W_{\rm het.}|_{T=0} =&& L^{10} (4\pi^2 \alpha^{\prime})^{-5}
\int_{\cal F} {{d^2 \tau}\over{4\tau_2^2}}
  (2 \pi \tau_2 )^{-5} \cdot
   {{1}\over{|\eta(\tau)|^{16}}} \cdot \left [
   \left ( {{{\bar{\Theta}}_3}\over{\bar{\eta}}}\right )^4 - \left ( {{{\bar{\Theta}}_4
   }\over{\bar{\eta}}}\right )^4 - \left ( {{ {\bar{\Theta}}_2}\over{\bar{\eta}}}\right )^4
   \right ]
   \cr
   && \quad\quad\quad\quad  \times
\left [ \left ({{\Theta_3}\over{\eta}} \right )^8 + \left (
{{\Theta_4}\over{\eta}} \right )^8 + \left
({{\Theta_2}\over{\eta}} \right )^8 \right ]^2 \quad .
\label{eq:holos}
\end{eqnarray}
Thus, $Z_{\rm het.}(\beta)$ describes the thermal mass spectrum of
$E_8$$\times$$E_8$ strings.

\vskip 0.1in It turns out that the desired result already exists
in the heterotic string literature. The modular invariant
possibilities for the sum over spin structures in the 10d
heterotic string have been classified, both by free fermion and by
orbifold techniques \cite{sw,dh,agmv,klt}, and there is a {\em
unique} nonsupersymmetric and tachyon-free solution with gauge
group $SO(16)$$\times$$SO(16)$. Recall the radius-dependent Wilson
line background described by Ginsparg in \cite{ginine} which
provides the smooth interpolation between the heterotic
$E_8$$\times$$E_8$ and $SO(32)$ theories in nine dimensions. We
have: ${\bf A}$$=$${{2}\over{x}}(1,0^7,-1,0^7)$, $x$$=$$
({{2}\over{\alpha^{\prime}}})^{1/2} r_{\rm circ.}$. Introducing
this background connects smoothly the 9D $SO(16)$$\times$$SO(16)$
string at generic radii with the supersymmetric 10d limit where
the gauge group is enhanced to $E_8$$\times$$E_8$. Note that the
states in the spinor lattices of $SO(16)$$\times$$SO(16)$
correspond to massless vector bosons only in the noncompact limit.
Generically, the $(17,1)$-dimensional heterotic momentum lattice
takes the form $E_8$$\oplus$$E_8$$\oplus$$U$. Here, $U$ is the
$(1,1)$ momentum lattice corresponding to compactification on a
circle of radius $r_{\rm circ.}$$=$$x(\alpha^{\prime}/2)^{1/2}$. A
generic Wilson line corresponds to a lattice boost as follows
\cite{ginine}:
\begin{equation}
({\bf p}; l_L, l_R) \to ({\bf p}'; l^{\prime}_L, l^{\prime}_R) =
({\bf p} + w x {\bf A}; u_L - {\bf p}\cdot {\bf A} -
{{wx}\over{2}} {\bf A} \cdot {\bf A} , u_R - {\bf p}\cdot {\bf A}
- {{wx}\over{2}} {\bf A} \cdot {\bf A} ) \quad . \label{eq:boost}
\end{equation}
${\bf p}$ is a 16-dimensional lattice vector in
$E_8$$\oplus$$E_8$. As shown in \cite{ginine}, the vacuum
functional of the supersymmetric 9d heterotic string, with generic
radius {\em and} generic Wilson line in the compact spatial
direction, can be written in terms of a sum over vectors in the
boosted lattice:
\begin{eqnarray}
W_{\rm SS} (r_{\rm circ.} ; {\bf A}) =&& L^{10}(4\pi^2
\alpha^{\prime})^{-5} \int_{\cal F} {{d^2 \tau}\over{4\tau_2^2}}
  (2 \pi \tau_2 )^{-5} |\eta(\tau)|^{-16}
{{1}\over{8}}  {{1}\over{{\bar{\eta}}^4}}
   ({\bar{\Theta}}_3^4 - {\bar{\Theta}}_4^4 - {\bar{\Theta}}_2^4 )
   \cr
   && \quad\quad\quad \times
\left [ {{1}\over{\eta^{16}}} \sum_{({\bf p}'; l^{\prime}_L ,
l^{\prime}_R)} q^{\half ({\bf p}^{\prime 2} +
  l_L^{\prime 2})} {\bar q}^{\half l_R^{\prime 2}}
\right ] \quad .   \label{eq:hollat}
\end{eqnarray}
$W_{\rm SS}$ describes the supersymmetric heterotic string with
gauge group $SO(16)$$\times$$SO(16)$ at generic radius. The
partition function of the nonsupersymmetric but tachyon-free 9d
string with gauge group $SO(16)$$\times$$SO(16)$ at generic radii
is given by \cite{dh,itoyama,sw,agmv,klt,gv}):
\begin{equation}
Z_{\rm NS} (r_{\rm circ.})= {{1}\over{4}} \left [
({{\Theta_2}\over{\eta}})^8 ({{\Theta_4}\over{\eta}})^8
({{{\bar{\Theta_3}}}\over{\eta}})^4 - ({{\Theta_2}\over{\eta}})^8
({{\Theta_3}\over{\eta}})^8 ({{{\bar{\Theta_4}}}\over{\eta}})^4 -
({{\Theta_3}\over{\eta}})^8 ({{\Theta_4}\over{\eta}})^8
({{{\bar{\Theta_2}}}\over{\eta}})^4 \right ]
  \sum_{n,w} q^{\half {\bf l}_L^2 } {\bar{q}}^{\half {\bf l}_R^2}
\quad . \label{eq:dualh}
\end{equation}
However, by identifying an appropriate interpolating function as
in previous sections and appropriate background field, we can
continuously connect this background to the supersymmetric
$E_8$$\times$$E_8$ string.

\vskip 0.1in Since $x$$=$$ ({{2}\over{\alpha^{\prime}}})^{1/2}
{{\beta}\over{2\pi}}$, from the viewpoint of the low-energy finite
temperature gauge theory the timelike Wilson line is simply
understood as imposing a modified axial gauge condition:
$A^0$$=$${\rm const}$. The dependence of the constant on
background temperature has been chosen to provide a shift in the
mass formula that precisely cancels the contribution from low
temperature $(n,0)$ tachyonic modes. As before, we begin by
identifying an appropriate modular invariant interpolating
function:
\begin{eqnarray}
 Z_{\rm het.} =&& {{1}\over{2}}
\sum_{n,w}
 \left [ ({{\Theta_2}\over{\eta}})^8
({{\Theta_4}\over{\eta}})^8 ({{{\bar{\Theta_3}}}\over{\eta}})^4 -
({{\Theta_2}\over{\eta}})^8 ({{\Theta_3}\over{\eta}})^8
({{{\bar{\Theta_4}}}\over{\eta}})^4 - ({{\Theta_3}\over{\eta}})^8
({{\Theta_4}\over{\eta}})^8 ({{{\bar{\Theta_2}}}\over{\eta}})^4
\right ]
  q^{\half {\bf l}_L^2 } {\bar{q}}^{\half {\bf l}_R^2}
\cr &&   {{1}\over{4}} \sum_{n,w} e^{\pi i (n+2w)}\left [ ( {{
{\bar{\Theta_3}} }\over{\eta}} )^4
-({{{\bar{\Theta_2}}}\over{\eta}})^4-({{{\bar{\Theta_4}}}\over{\eta}})^4
\right ] \left [ ({{\Theta_3}\over{\eta}})^{16} + (
{{\Theta_4}\over{\eta}})^{16} + ({{\Theta_2}\over{\eta}})^{16})
\right ] q^{\half {\bf l}_L^2 } {\bar{q}}^{\half {\bf l}_R^2} \cr
\quad && \quad + {{1}\over{2}} \sum_{n,w} e^{\pi i (n+2w)}
 \{ ({{{\bar{\Theta_3}}}\over{\eta}})^4 ({{\Theta_3}\over{\eta}})^8
 \left [
({{\Theta_4}\over{\eta}})^8 + ({{\Theta_2}\over{\eta}})^8 \right ]
- ({{{\bar{\Theta_2}}}\over{\eta}})^4 ({{\Theta_2}\over{\eta}})^8
\left [ ({{\Theta_3}\over{\eta}})^8 + ({{\Theta_4}\over{\eta}})^8
\right ] \cr && \quad \quad \quad -
({{{\bar{\Theta_4}}}\over{\eta}})^4 ({{\Theta_4}\over{\eta}})^8
\left [ ({{\Theta_2}\over{\eta}})^8 + ({{\Theta_3}\over{\eta}})^8
\right ] \}
  q^{\half {\bf l}_L^2 } {\bar{q}}^{\half {\bf l}_R^2} \quad .
\label{eq:iden}
\end{eqnarray}
As in previous sections, the first term within square brackets has
been chosen as the nonsupersymmetric sum over spin structures for
a {\em chiral} type 0 string. This function appears in the sum
over spin structures for the tachyon-free $SO(16)$$\times$$SO(16)$
string given above. Notice that taking the $x$$\to$$\infty$ limit,
by similar manipulations as in the type II case, yields the
partition function of the supersymmetric 10D $E_8$$\times$$E_8$
string.

\vskip 0.1in Consider accompanying the $SO(17,1)$ transformation
described above with a lattice boost that decreases the size of
the interval \cite{ginine}:
\begin{equation}
 e^{-\alpha_{00}} = {{1}\over{1+|{\bf A}|^2/4}} \quad .
\label{eq:scale}
\end{equation}
This recovers the Spin(32)/Z$_2$ theory compactified on an
interval of size $2/x$, but with Wilson line ${\bf A}$$=$$x{\rm
diag}(1^8,0^8)$ \cite{ginine}. Thus, taking the large radius limit
in the {\em dual} variable, and with dual Wilson line background,
yields instead the spacetime supersymmetric 10D $Spin(32)/Z_2$
heterotic string. It follows that the $E_8$$\times$$E_8$ and
$SO(32)$ heterotic strings share the same tachyon-free finite
temperature ground state with gauge symmetry
$SO(16)$$\times$$SO(16)$. The Kosterlitz-Thouless transformation
at $T_C$$=$$1/2\pi\alpha^{\prime 1/2}$ is a self-dual continuous
phase transition in this theory.

\vskip 0.1in The thermal duality transition in this theory is in
the universality class of the Kosterlitz-Thouless transition
 \cite{kt}: namely, the partial
derivatives of the free energy to arbitrary order are analytic
functions of temperature. The duality transition interchanges
thermal momentum modes of the $E_8$$\times$$E_8$ theory with
winding modes of the $Spin(32)/{\rm Z}_2$ theory, and vice versa.
Note that the vacuum functional, the Helmholtz and Gibbs free
energies, the internal energy, and all subsequent thermodynamic
potentials, are both finite and tachyon-free.

\subsection{Holography and the Duality Phase Transition}

\vskip 0.1in In the closed bosonic string theory, the generating
functional for connected one-loop vacuum string graphs is
invariant under the thermal duality transformation: $W(T)$ $=$
$W(T^2_c/T)$, with self-dual temperature, $T_c$ $=$ $1/2 \pi
\alpha^{\prime 1/2}$. As already noted by Polchinski
\cite{polchinskibook}, we can infer the following thermal duality
relation which holds for both the Helmholtz free energy, $F(T)$
$=$ $-T \cdot W(T)$, and the effective potential, $\rho(T)$ $=$
$-T \cdot W(T)/V$ of the closed bosonic string:
\begin{equation}
F(T)  = {{T^2}\over{T_C^2}} F({{T_C^2}\over{T}}) , \quad \quad
\rho(T) = {{T^2}\over{T^2_C }} \rho({{T_C^2}\over{T}}) \quad .
\label{eq:thermi}
\end{equation}
In the case of the heterotic string, the thermal duality relation
instead relates, respectively, the free energies of the
$E_8$$\times$$E_8$ and ${\rm Spin}(32)/{\rm Z_2}$ theories:
\begin{equation}
F(T)_{E_8\times E_8}  = {{T^2}\over{T_C^2}}
F({{T_C^2}\over{T}})_{{\rm Spin}(32)/{\rm Z}_2} , \quad \quad
\rho(T)_{E_8 \times E_8}  = {{T^2}\over{T^2_C }}
\rho({{T_C^2}\over{T}})_{\rm Spin (32)/Z_2} \quad .
\label{eq:thermih}
\end{equation}
 Consider the high temperature limit of this
expression:
\begin{equation}
\lim_{T \to \infty } \rho(T)_{E_8 \times E_8} = \lim_{T \to
\infty} {{T^2}\over{T^2_C }} \rho({{T_C^2}\over{T}})_{\rm Spin
(32)/Z_2}
 =  \lim_{(T_C^2/T) \to 0} {{T^2}\over{T^2_C }}
\rho({{T_C^2}\over{T}})_{\rm Spin(32)/Z_2}
 =  {{T^2}\over{T^2_C }} \rho(0)_{\rm Spin(32)/Z_2}
\quad , \label{eq:thermasy}
\end{equation}
where $\rho(0)$ is the cosmological constant, or vacuum energy
density, at zero temperature. Note that it is finite. Thus, at
high temperatures, the free energy of either heterotic theory
grows as $T^2$. In other words, the growth in the number of
degrees of freedom at high temperature in the heterotic string
ensemble is only as fast as in a {\em two-dimensional} field
theory. This is significantly slower than the $T^{10}$ growth of
the high temperature degrees of freedom expected in the
ten-dimensional low energy field theory.

\vskip 0.1in Notice that the prefactor, $\rho(0)/T_C^2$, in the
high temperature relation is unambiguous, a consequence of the
normalizability of the generating functional of one-loop vacuum
graphs in string theory \cite{poltorus}. It is also background
dependent: it is computible as a continuously varying function of
the background fields upon compactification to lower spacetime
dimension \cite{gv}. The relation in Eq.\ (\ref{eq:thermasy}) is
unambiguous evidence of the holographic nature of perturbative
heterotic string theory: {\em there is a drastic reduction in the
degrees of freedom in string theory at high temperature, a
conjecture first made in \cite{aw}}.

\vskip 0.1in Starting with the duality invariant expression for
the string effective action functional, we can derive the
thermodynamic potentials of the heterotic string ensemble. The
Helmholtz free energy follows from the definition below Eq.\
(\ref{eq:het}), and is clearly finite at all temperatures, with no
evidence for either divergence or discontinuity. The internal
energy of the heterotic ensemble takes the form:
\begin{eqnarray}
 U \equiv&& - \left ( {{\partial W}\over{\partial \beta }} \right )_V
 \cr
 =&&
 L^9 (4\pi^2 \alpha^{\prime})^{-9/2} \half \int_{\cal F}
{{|d\tau|^2}\over{4\tau_2^2}} (2\pi\tau_2)^{-9/2}
   |\eta(\tau)|^{-16}
{{4\pi \tau_2}\over{\beta}} \sum_{n,w \in {\rm Z} }
  \left ( w^2 x^2 - {{n^2}\over{x^2}}  \right )
\cdot q^{ {{1}\over{2}}{\bf l}_L^2 }
     {\bar q}^{{{1}\over{2}}{\bf l}_R^2 }
\cdot Z_{\rm [SO(16)]^2}  , \cr && \label{eq:term}
\end{eqnarray}
where $Z_{\rm [SO(16)]^2}$ denotes the sums over spin structures
appearing in Eq.\ (\ref{eq:iden}). Notice that $U(\beta)$ vanishes
precisely at the self-dual temperature,
$T_c$$=$$1/2\pi\alpha^{\prime 1/2}$, $x_c$$=$${\sqrt{2}}$, where
the internal energy contributed by winding sectors cancels that
contributed by momentum sectors. Note also that the internal
energy changes sign at $T$ $=$ $T_C$. Hints of this behaviour are
already apparent in the numerical analyses of the one-loop
effective potential given in \cite{gv,itoyama}.

\vskip 0.1in It is easy to demonstrate the analyticity of
infinitely many thermodynamic potentials in the vicinity of the
critical point. It is convenient to define:
\begin{eqnarray}
[d \tau ] &&\equiv \half L^9 (4\pi^2 \alpha^{\prime})^{-9/2} \left
[ {{|d\tau|^2}\over{4\pi\tau_2^2}} (2\pi\tau_2)^{-9/2}
   |\eta(\tau)|^{-16}
\cdot Z_{\rm [SO(16)]^2} e^{2\pi i nw \tau_1} \right ] \cr
y(\tau_2;x)
   &&\equiv 2\pi \tau_2 \left ( {{n^2}\over{x^2}} + w^2 x^2
  \right ) \quad .
\label{eq:vars}
\end{eqnarray}
Denoting the $m$th partial derivative with respect to $\beta$ at
fixed volume by $W_{(m)}$, $y_{(m)}$, we note that the higher
derivatives of the vacuum functional take the simple form:
\begin{eqnarray}
W_{(1)} =&& \sum_{n,w} \int_{\cal F} [d\tau] e^{-y} (-y_{(1)})
%, \quad
\cr W_{(2)} =&&
 \sum_{n,w} \int_{\cal F} [d\tau] e^{-y}
(-y_{(2)} + (-y_{(1)})^2 ) \cr
%, \quad
W_{(3)} =&& \sum_{n,w} \int_{\cal F} [d\tau] e^{-y} (-y_{(3)} -
y_{(1)} y_{(2)} +  (-y_{(1)})^3 ) \cr
 \cdots =&& \cdots
\cr W_{(m)} =&& \sum_{n,w} \int_{\cal F} [d\tau] e^{-y} (-y_{(m)}
- \cdots +  (-y_{(1)})^m ) \quad . \label{eq:effm}
%\end{equation}
\end{eqnarray}
Referring back to the definition of $y$, it is easy to see that
both the vacuum functional and, consequently, the full set of
thermodynamic potentials, are analytic in $x$. Notice also that
third and higher derivatives of $y$ are determined by the momentum
modes alone:
\begin{equation}
y_{(m)} = (-1)^m n^2 {{(m+1)! }\over{x^{m+2}}} , \quad m \ge 3
\quad . \label{eq:ders}
\end{equation}
For completeness, we give explicit results for the first few
thermodynamic potentials:
\begin{equation}
F = - {{1}\over{\beta}} W_{(0)} , \quad U = - W_{(1)} , \quad S =
W_{(0)} - \beta W_{(1)} , \quad C_V = \beta^2 W_{(2)} , \cdots
\quad . \label{eq:thermodl}
\end{equation}
The entropy is given by the expression:
\begin{equation}
S = \sum_{n,w} \int_{\cal F} [d\tau] e^{-y}
  \left [ 1 + 4\pi \tau_2 ( - {{n^2}\over{x^2}} + w^2 x^2 )
\right ] \quad , \label{eq:entropy1}
\end{equation}
For the specific heat at constant volume, we have:
\begin{equation}
C_V =  \sum_{n,w} \int_{\cal F} [d\tau] e^{-y}
  \left [   16 \pi^2 \tau_2^2 ( - {{n^2}\over{x^2}} + w^2 x^2 )^2
   - 4\pi \tau_2( 3 {{n^2}\over{x^2}} + w^2 x^2 )
\right ] . \label{eq:spc}
\end{equation}
Explicitly, the Helmholtz free energy takes the form:
\begin{equation}
F (\beta) = - \half {{1}\over{\beta}} L^9 (4\pi^2
\alpha^{\prime})^{-9/2}  \int_{\cal F}
{{|d\tau|^2}\over{4\pi\tau_2^2}} (2\pi\tau_2)^{-9/2}
   |\eta(\tau)|^{-16}
\left [ \sum_{n,w} Z_{\rm [SO(16)]^2} q^{ {{1}\over{2}}{{\alpha'
p_L^2}\over{2}} }
    {\bar q}^{{{1}\over{2}} {{\alpha' p_R^2}\over{2}}}
\right ] \quad , \label{eq:freeee}
\end{equation}
while for the entropy of the heterotic string ensemble, we have
the result:
\begin{eqnarray}
S (\beta) =&& \half L^9 (4\pi^2 \alpha^{\prime})^{-9/2} \int_{\cal
F} {{|d\tau|^2}\over{4\pi\tau_2^2}} {{(2\pi\tau_2)^{-9/2}}\over{
   |\eta(\tau)|^{16}}}
\sum_{n,w} \left [
   1 + 4 \pi \tau_2 ( - {{n^2}\over{x^2}} + w^2 x^2 )
\right ] Z_{\rm [SO(16)]^2} q^{ {{1}\over{2}}{{\alpha'
p_L^2}\over{2}} }
    {\bar q}^{{{1}\over{2}} {{\alpha' p_R^2}\over{2}}} .
    \cr &&
\label{eq:entropy}
\end{eqnarray}
The thermodynamic potentials of the heterotic ensemble are finite
normalizable functions at all temperatures starting from zero. In
summary, the heterotic string ensemble displays a continuous phase
transition at the self-dual temperature, mapping thermal winding
modes of the $E_8$$\times$$E_8$ theory to thermal momentum modes
of the ${\rm Spin}(32)/{\rm Z}_2$ theory, and vice versa,
unambiguously identifying a phase transition of the
Kosterlitz-Thouless type \cite{kt,bosonic}.

\section{Type I and Type II Open and Closed String Theories}

The Type I and Type II string theories can have both open and
closed string sectors, and the vacuum can contain Dbranes: sources
for Ramond-Ramond charge, with worldvolume Yang-Mills fields
\cite{polchinskibook,zeta}. It is therefore helpful to consider
them in a unified treatment. We will begin with the pure oriented
closed string sector common to all of these theories.

\subsection{Closed Oriented Superstring Sector}

\vskip 0.1in We begin with a discussion of the pure type II
oriented closed string one-loop vacuum functional. A thermal
duality transformation mapping the IIA string to the IIB string
simply maps IIA winding to IIB momentum modes, and vice versa. In
the absence of a Ramond-Ramond sector, the expression for the
normalized generating functional of connected one-loop vacuum
graphs takes the form:
\begin{equation}
W_{\rm II} = L^{9} (4\pi^2 \alpha^{\prime})^{-9/2} \int_{\cal F}
{{d^2 \tau}\over{4\tau_2^2}}
  (2 \pi \tau_2 )^{-4} |\eta(\tau)|^{-14} Z_{\rm II} (\beta)
\quad ,
\label{eq:typeII}
\end{equation}
where the spatial volume $V$$=$$L^9$, while the dimensionless
(scaled) spatial volume is $ L^{9} (4\pi^2 \alpha^{\prime})^{-9/2}
$. The inverse temperature is given by $\beta $ $=$ $ 2 \pi r_{\rm
circ.}$. Notice that in the $\alpha^{\prime}$ $\to$ $0$ limit one
can simultaneously take the size of the \lq\lq box" to infinity
while keeping the rescaled volume fixed. This defines the approach
to the thermodynamic limit. The function ${\rm Z}_{\rm II~
orb.}(\beta)$ is the product of contributions from worldsheet
fermions and bosons, ${\rm Z}_F Z_{\rm B}$, and is required to
smoothly interpolate between finite temperature and the spacetime
supersymmetric zero temperature limit. The spectrum of thermal
modes will be unambiguously determined by modular invariance. The
spacetime supersymmetry breaking projection, $(-1)^{N_F}$, is
modified by the introduction of {\em phases} in the interpolating
function. Such phases can depend on thermal mode number. They must
be chosen {\em compatible with the requirement that spacetime
supersymmetry is restored in the zero temperature limit of the
interpolating function}. We comment that temperature dependent
phases in the free enerrgy were first proposed by Atick and Witten
in \cite{aw}. The {\em unique} modular invariant interpolating
function satisfying these requirements is:
\begin{eqnarray}
Z_{\rm II} (\beta) =&& {{1}\over{2}} {{1}\over{\eta{\bar{\eta}}}}
\left [ \sum_{w,n \in {\rm Z} }
   q^{{{1}\over{2}} ({{n }\over{x}} + wx )^2 }
     {\bar q}^{{{1}\over{2}} ({{n}\over{x}} - wx )^2 }
 \right ]
\{ (|\Theta_3 |^8
  + |\Theta_4|^8  + |\Theta_2|^8 ) \cr
  \quad&&\quad\quad + e^{ \pi i (n + 2w ) }
 [ (\Theta_2^4 {\bar{\Theta}}^4_4 + \Theta_4^4 {\bar{\Theta}}_2^4)
  - (\Theta_3^4 {\bar{\Theta}}_4^4 + \Theta_4^4 {\bar{\Theta}}_3^4
   + \Theta_3^4 {\bar{\Theta}}_2^4 + \Theta_2^4 {\bar{\Theta}}_3^4 ) ]
\quad ,
\label{eq:bosod}
\end{eqnarray}
where $x$$=$$ ({{2}\over{\alpha^{\prime}}})^{1/2} r_{\rm circ.}$.
The world-sheet fermions have been conveniently complexified into
left- and right-moving Weyl fermions. As in the superstring, the
spin structures for all ten left- and right-moving fermions,
$\psi^{\mu}$, ${\bar{\psi}}^{\mu}$, $\mu$ $=$ $0$, $\cdots$, $9$,
are determined by those for the world-sheet gravitino associated
with left- and right-moving N=1 superconformal invariances.

\vskip 0.1in
To understand our result for the correct interpolating function, first recall
the expression for ${\rm Z}_{SS}$--- the zero temperature, spacetime
supersymmetric, limit of our function given by the ordinary GSO projection:
\begin{equation}
 Z_{\rm SS} =
     {{1}\over{4}}  {{1}\over{\eta^4{\bar{\eta}}^4}}
  \left [ (\Theta_3^4 - \Theta_4^4 - \Theta_2^4)
   ({\bar{\Theta}}_3^4 - {\bar{\Theta}}_4^4 - {\bar{\Theta}}_2^4 )
\right ]
 \quad ,
\label{eq:IIs}
\end{equation}
Notice that the first of the relative signs
in each round bracket preserves the tachyon-free condition. The second relative sign
determines whether spacetime supersymmetry is preserved in the zero
temperature spectrum. Next, notice that $Z_{SS}$ can be rewritten
using theta function identities as follows:
\begin{eqnarray}
 Z_{\rm SS } &&=
\{ [ |\Theta_3 |^8
  + |\Theta_4|^8  + |\Theta_2|^8 ] \cr
  \quad&&\quad\quad +
 [ (\Theta_2^4 {\bar{\Theta}}^4_4 + \Theta_4^4 {\bar{\Theta}}_2^4)
  - (\Theta_3^4 {\bar{\Theta}}_4^4 + \Theta_4^4 {\bar{\Theta}}_3^4
   + \Theta_3^4 {\bar{\Theta}}_2^4 + \Theta_2^4 {\bar{\Theta}}_3^4 ) ]
  \}
 \quad .
\label{eq:IIid}
\end{eqnarray}
Either of the two expressions within square
brackets is modular invariant.
The first may be recognized as the nonsupersymmetric sum over
spin structures for the type 0 string \cite{polchinskibook}.

\vskip 0.1in
Thus, the interpolating function captures the desired zero
temperature limits of both the IIA and IIB strings: for
large $\beta_{IIA}$, terms with $w_{IIA}$$\neq$$0$
decouple in the double summation since they are exponentially
damped. The remaining terms are resummed by a Poisson
resummation, thereby inverting the $\beta_{IIA}$ dependence in the
exponent, and absorbing the phase factor in a shift of argument.
The large $\beta_{IIA}$ limit can then be taken smoothly
providing the usual momentum integration for a non-compact dimension,
plus a factor of $\tau_2^{-1/2}$ from the Poisson resummation.
This recovers the supersymmetric zero temperature partition function.

\vskip 0.1in
As regards thermal duality, small $\beta_{IIA}$ maps to large
$\beta_{IIB}$$=$$\beta_C^2/\beta_{IIA}$, also interchanging the
identification of momentum and winding modes,
$(n,w)_{IIA}$$\to$$(n'=w,w'=n)_{IIB}$.
We can analyze the limit of large dual
inverse temperature as before, obtaining the zero temperature
limit of the dual IIB theory.
At any intermediate temperature, all of the
thermal modes contribute to the vacuum functional with
a phase that takes values $(\pm 1)$ only. Note that the spacetime
fermions of the zero temperature spectrum now contribute with
a reversed phase, evident in the first term in
Eq.\ (\ref{eq:bosod}), as required by the
thermal boundary conditions.
In summary, the vacuum functional of the IIA string
maps precisely into the vacuum functional of the IIB string
under a thermal duality transformation.

\subsection{Tachyonic Thermal Momentum and Thermal Winding Modes}

\vskip 0.1in We now point out a peculiarity of the type IIA and
IIB canonical ensemble at finite temperature. First, recall some
pertinent facts from quantum field theory. It is generally assumed
to be the case \cite{kirzhlinde} that in a field theory known to
be perturbatively renormalizable at zero temperature, the new
infrared divergences introduced by a small variation in the
background temperature can be self-consistently regulated by a
suitable extension of the renormalization conditions, at the cost
of introducing a finite number of additional counter-terms. The
zero temperature renormalization conditions on 1PI Greens
functions are conveniently applied at zero momentum, or at fixed
spacelike momentum in the case of massless fields. Consider a
theory with one or more scalar fields. Then the renormalization
conditions must be supplemented by stability constraints on the
finite temperature effective potential \cite{kirzhlinde}:
\begin{equation}
{{\partial V_{\rm eff.}(\phi_{\rm cl.})}\over{\partial
\phi_{\rm cl.}}} = 0, \quad
\quad
{{\partial^2 V_{\rm eff.}(\phi_{\rm cl. })}\over{\partial \phi_{\rm cl.}^2 }} \equiv
G_{\rm \phi}^{-1} (k)|_{k=0}
\geq 0 \quad .
\label{eq:stab}
\end{equation}
Here, $V_{\rm eff.}(\phi_{\rm cl.})$$=$$-\Gamma(\phi_{\rm cl.})/V\beta$, where
$\Gamma$ is the effective action functional, or sum of
connected 1PI vacuum diagrams at finite temperature. In the absence of
nonlinear field configurations, and for perturbation theory in a small coupling
about the free field vacuum, $|0>$, $\phi_{\rm cl.}$ is simply the
expectation value of the scalar field: $\phi_{\rm cl.}$$=$$<0 | \phi(x) | 0 >$.
The first condition holds in the absence of an external source at every
extremum of the effective potential.
The second condition states that the renormalized masses of physical fields
must not be driven to imaginary values at any non-pathological and stable
minimum of the effective potential.

\vskip 0.1in The conditions in Eq.\ (\ref{eq:stab}) are, in fact,
rather familiar to string theorists. Weakly coupled superstring
theories at zero temperature are replete with scalar fields and
their vacuum expectation values, or moduli, parameterize a
multi-dimensional space of degenerate vacua. Consider the effect
of an infinitesimal variation in the background temperature. Such
an effect will necessarily break supersymmetry, and it is
well-known that in the presence of a small spontaneous breaking of
supersymmetry the dilaton potential will generically develop a
runaway direction, signalling an instability of precisely the kind
forbidden by the conditions that must be met by a non-pathological
ground state \cite{dinesei}. This is worrisome. Namely, while the
zero temperature effective potential may be correctly minimized
with respect to the renormalizable couplings in the potential and,
in fact, vanishes in a spacetime supersymmetric ground state, one
or more of the scalar masses could be driven to imaginary values
in the presence of an {\em infinitesimal} variation in background
temperature. Such a quantum field theory would be simply
unacceptable both as a self-consistent effective field theory in
the Wilsonian sense, and also as a phenomenological model for a
physical system \cite{kirzhlinde}. The same conclusion would hold
for any weakly coupled superstring theory with these pathological
properties. Fortunately, as is convincingly demonstrated in the
follow-up work \cite{relevant}, the type IIA and type IIB thermal
ensembles we have described in the previous sub-section with
trivial Ramond-Ramond sector are indeed unstable. Most
importantly, the flow of the worldsheet renormalization group (RG)
is in the direction {\em towards} the supersymmetric zero
temperature vacuum of the type IIA, or type IIB, string theory.

\vskip 0.1in Let us explain why the type IIA or type IIB thermal
ensemble is pathological in the absence of a Ramond-Ramond sector,
or of Yang-Mills gauge fields. To check for potential tachyonic
instabilities in the expressions given in Eq.\ (\ref{eq:bosod}),
consider the mass formula in the (NS,NS) sector for world-sheet
fermions, with ${\bf l}_L^2 $ $=$ $ {\bf l}_R^2$, and
$N$$=$${\bar{N}}$$=$$0$:
\begin{equation}
({\rm mass})^2_{nw} =  {{2}\over{\alpha^{\prime}}}
 \left [ - 1 +
{{2 \alpha^{\prime}\pi^2 n^2 }\over{\beta^2}} +
{{\beta^2w^2}\over{8\pi^2 \alpha^{\prime} }}  \right ] \quad .
\label{eq:massII}
\end{equation}
This is the only sector that contributes tachyons to the thermal
spectrum. Recall that $\beta_C^2$$=$$4\pi^2 \alpha^{\prime}$ is
the self-dual point. A nice check is that the momentum mode number
dependence contained in the phase factors does not impact the
spacetime spin-statistics relation in the NS-NS sector: all
potentially tachyonic states are spacetime scalars as expected.
Notice that the $n$$=$$w$$=$$0$ sector common to both type II
string theories contains a potentially tachyonic state whose mass
is now temperature dependent. However, it corresponds to an
unphysical tachyon. The potentially tachyonic physical states are
the pure momentum and pure winding states, $(n,0)$ and $(0,w)$,
with $N$$=$${\bar{N}}$$=$$0$. As in the closed bosonic string
analysis, we can compute the temperatures below, and beyond, which
these modes become tachyonic in the absence of oscillator
excitations. Each momentum mode, $(\pm n, 0)$, is tachyonic {\em
upto} some critical temperature, $T^2_n$ $=$ $1/2 n^2\pi^2
\alpha^{\prime} $, after which it turns marginal (massless).
Conversely, each winding mode $(0,\pm w)$, is tachyonic {\em
beyond} some critical temperature, $T^2_w$ $=$ $w^2/8\pi^2
\alpha^{\prime}$. {\em The thermal spectrum of pure type IIA or
IIB closed oriented strings is unstable at all temperatures
starting from zero.}

\vskip 0.1in The source of the thermal instability discussed here
should not be confused with the gravitational instability of flat
spacetime found in the finite temperature field theory analysis of
\cite{gyp}. Nor is it to be confused with the famous Jeans
instability of gravitating matter in the limit of infinite spatial
volume \cite{aw}. Our considerations have been limited to the
internal consistency conditions of the {\em free} closed string
thermal spectrum. In particular, the ensemble of closed strings
does not gravitate at this order in perturbation theory.

\vskip 0.1in We should now remined the reader that {\em in the
presence of Dbranes}, the type IIA and IIB strings acquire an open
string sector and, consequently, nonabelian gauge fields in the
massless spectrum. Dbranes are BPS sources, and half of the
supersymmetries of the type II theory are broken in their
presence. As we will show in section 3, in the type I$^{\prime}$
theory with D8branes, the presence of nonabelian gauge fields
enables invoking a Wilson line background. In the presence of a
{\em temperature} dependent (timelike) Wilson line, it is a simple
matter to remove all of the thermal tachyons from the spectrum of
the type II theory. This mechanism has a precise analog in the
case of both the closed heterotic string ensemble, as well as in
the type I$^{\prime}$ ensemble of open and closed strings as
demonstrated in the following sections.

\section{Canonical Ensemble of Type I$^{\prime}$ Strings}

\subsection{Cancellation of Ramond-Ramond Scalar Tadpole}

\vskip 0.1in In the previous section, we have seen that the two
distinct ten-dimensional supersymmetric $E_8$$\times$$E_8$ and
${\rm Spin(32)/Z_2}$ heterotic string theories have a common
tachyon-free ground state at finite temperature in the presence of
a temperature dependent timelike Wilson line. The gauge group at
finite temperatures is $SO(16)$$\times$$SO(16)$. Let us now
consider what happens to the type I and type I$^{\prime}$ open and
closed string theories at finite temperature.

\vskip 0.1in It will be convenient to begin with the timelike
T-dual type I$^{\prime}$ string theory with 16 D8branes on each of
two orientifold planes, separated by $\beta$ in the Euclidean time
direction \cite{polchinskibook,decon}. From the perspective of the
original type IB string theory with gauge group $O(32)$, we have
turned on a temperature-dependent timelike Wilson line,
$A_0$$=$${{1}\over{\beta}}((1)^8,0^8)$, in the worldvolume of the
space-filling D9branes. This breaks the gauge symmetry to
$O(16)$$\times$$O(16)$. In the type I$^{\prime}$ picture, the
counting of zero length open strings connecting a pair of D8branes
at finite temperature is as follows: we have a total of
($2$$\cdot$$8$$\cdot$$7$)$\cdot$$2$ massless states from zero
length open strings connecting a pair of D8branes on the same
orientifold plane. In addition, there are $8$$\cdot$$2$ photons
from zero length open strings connecting each D8brane to itself.
Open strings connecting branes on distinct orientifold planes are
ordinarily massive, resulting in a breaking of the gauge group
from $O(32)$ to $O(16)$$\times$$O(16)$ at finite temperature.

\vskip 0.1in Consider the free energy, $F(\beta)$, of a gas of
free type I$^{\prime}$ strings in this ground state. $F$ is
obtained from the generating functional for connected vacuum
string graphs, $F(\beta)$$=$$- W(\beta)/\beta$. The one-loop free
energy receives contributions from worldsurfaces of four different
topologies \cite{polchinskibook,zeta}: torus, annulus, Mobius
Strip, and Klein Bottle. The torus is the sum over closed oriented
worldsheets and the result is, therefore, identical to that
derived in section 2 for the type IIB string theory, refer to
Eqs.\ (\ref{eq:typeII}) and (\ref{eq:bosod}). Notice that the
closed oriented string sector of the type IB theory does not
distinguish between the Dbrane worldvolume and the bulk spacetime
orthogonal to the branes, since the closed oriented strings live
in all ten dimensions of spacetime. Nor does this sector have any
knowledge about the Yang-Mills sector, or of the timelike Wilson
line. The comments we have made earlier regarding worldsheet
renormalization group (RG) flow in the direction {\em towards} the
noncompact supersymmetric vacuum apply for this sector of the
theory: although we begin with an, a priori, non-vanishing thermal
contribution to the vacuum energy, RG flow takes us back to the
supersymmetric infra-red stable fixed point. Thus, the torus
contribution to the vacuum energy vanishes, and supersymmetry is
{\em not} broken by thermal effects in this sector of the
unoriented type I open and closed string theory.

\vskip 0.1in The one-loop contribution to $W$ from the remaining
three worldsheet topologies is given by the Polyakov path integral
summing surfaces with two boundaries, with a boundary and a
crosscap, or with two crosscaps \cite{polchinskibook}. The
boundaries and crosscaps are localized on the orientifold plane
but the worldsheet itself can extend into the transverse Dirichlet
directions. $W(\beta)|_{\beta =\infty}$ is required to agree with
the vacuum functional of the supersymmetric $O(32)$ type
I$^{\prime}$ string at zero temperature. We also require a
tachyon-free thermal spectrum which retains the good ultraviolet
and infrared behavior of the supersymmetric zero temperature
limit. As in the case of the supersymmetric vacuum, consistency of
the finite temperature type I$^{\prime}$ string vacuum requires
cancellation of the tadpole for the unphysical Ramond-Ramond
scalar \cite{polchinskibook}. Such a scalar potential could appear
as the Poincare dual of an allowed ten-form potential in the
unoriented type I string theory, but notice that the associated
11-form field strength is {\em required} to be identically zero in
this ten-dimensional theory.

\vskip 0.1in On the other hand, the usual cancellation between
closed string NS-NS and R-R exchanges resulting from spacetime
supersymmetry must not hold except in the zero temperature limit.
This is in precise analogy with the finite temperature analysis
given earlier for closed string theories and is achieved by
introducing temperature dependent phases in the vacuum functional
\cite{aw}. We will insert an identical temperature-dependent phase
for the contributions to $F(\beta)$ from worldsheets with each of
the three remaining classes of one-loop graphs--- annulus, Mobius
Strip, and Klein Bottle, in order to preserve the form of the RR
scalar tadpole cancellation in the zero temperature ground state.
The result for the free energy at one-loop order, with $N$$=$$16$
D8branes on each orientifold plane, takes the form:
\begin{eqnarray}
F = && - \beta^{-1} L^9 (4\pi^2 \alpha^{\prime})^{-9/2}
\int_0^{\infty} {{dt}\over{8t}}e^{-\beta^2 t / 2 \pi
\alpha^{\prime}}
  {{(2\pi t )^{-9/2}}\over{
  \eta(it)^{8}}} \sum_{n \in {\rm Z} }
[ ~  2^{-8} N^2 ( Z_{\rm [0]} - e^{\i \pi n } Z_{\rm [1]}) \cr
\quad&& \quad\quad\quad  +   Z_{\rm [0]} - e^{\i \pi n } Z_{\rm
[1]} ~ - ~ 2^{-3} N  Z_{\rm  [0]}  + ~ 2^{-3} N e^{\i \pi n }
Z_{\rm [1]} ~ ] \times  q^{4 \alpha^{\prime}\pi^2 n^2 /\beta^2}
\label{eq:freeIp}
\end{eqnarray}
We emphasize that this expression is uniquely singled out by the
requirement of tadpole cancellation for the dualized RR scalar. It
also interpolates smoothly between the supersymmetric zero
temperature limit, with gauge group $O(32)$, and the finite
temperature result with gauge group $O(16)$$\times$$O(16)$. Notice
that the momentum modes have no winding mode counterparts in the
open string thermal spectrum because of the absence of thermal
duality. The Matsubara frequency spectrum is given by the
eigenvalues for timelike momentum: $p_0$$=$$2n\pi/\beta$, with
$n$$\in$${\rm Z}$. Finally, as an important consistency check,
notice that there are no new tadpoles in the expression for $W$
due to finite temperature excitations alone.

\vskip 0.1in The variable $q$$=$$e^{-2\pi t}$ is the modular
parameter for the cylinder, and we have used the identifications:
$2t_K$$=$$t_C$, and $2t_M$$=$$t_C$, to express the Mobius strip
and Klein bottle amplitudes in terms of the cylinder's modular
parameter. As in \cite{polchinskibook}, the subscripts $[0]$,
$[1]$, denote, respectively, NS-NS and R-R closed string
exchanges. Thus,
\begin{eqnarray}
Z_{\rm [0]} =&& ({{\Theta_{00}(it;0)}\over{\eta(it)}})^4 -
({{\Theta_{10}(it;0)}\over{\eta(it)}})^4
\cr
Z_{\rm [1]} =&& ({{\Theta_{01}(it;0)}\over{\eta(it)}})^4 -
({{\Theta_{11}(it;0)}\over{\eta(it)}})^4
\label{eq:relats}
\end{eqnarray}
where the (00), (10), (01), and (11), denote, respectively,
(NS-NS), (R-NS), (NS-R), and (R-R), boundary conditions on
worldsheet fermions in the closed string sector
\cite{polchinskibook}. We emphasize that, analogous to the
supersymmetric zero temperature limit, cancellation of the tadpole
for the unphysical Ramond-Ramond scalar {\em requires} $N$$=$$16$.
It can be verified that the dilaton tadpole is also absent as a
consequence of this constraint. Remarkably, we have found a
nonsupersymmetric type I vacuum with supersymmetry broken by
thermal effects, but without the appearance of a dilaton tadpole.

\vskip 0.2in
\subsection{Holography in the Open and Closed String Ensemble}

\vskip 0.25in The expression for the generating functional of
one-loop vacuum string graphs given in Eq.\ (\ref{eq:freeIp})
explicitly violates thermal duality and so, unlike the case of
closed string gases, there is no analogous thermal self-duality
relation holding at $T_C$. Nevertheless, we will find concrete
evidence of a high temperature holographic phase by direct
inspection of the modular integration. Notice that the high
temperature behavior of $F(\beta)$ can be unambiguously identified
because the dominant contribution in this limit comes from the
$t$$\to$$0$ regime. A modular transformation on the argument of
the theta functions, $t$$\to$$1/t$, puts the integrand in a
suitable form for term-by-term expansion in powers of $e^{-1/t}$;
this enables term-by-term evaluation of the modular integral.
Isolating the $t$$\to$$0$ asymptotics in the standard way
\cite{polchinskibook}, we obtain a most unexpected result:
\begin{eqnarray}
\lim_{\beta \to 0} F = && - \lim_{\beta \to 0} \beta^{-1}L^9
(4\pi^2 \alpha^{\prime})^{-9/2} \int_0^{\infty} {{dt}\over{8t}}
e^{ -\beta^2t /2\pi \alpha^{\prime} }
  {{(2\pi t )^{-1/2}}\over{
  \eta(i/t)^{8}}}\sum_{n \in {\rm Z} }
[ ~  2^{-8} N^2 ( Z_{\rm [0]} (i/t)- e^{\i \pi n } Z_{\rm [1]})
(i/t)
 \cr
\quad && \quad\quad +   Z_{\rm [0]} (i/t) - e^{\i \pi n } Z_{\rm
[1]} (i/t)
 ~ - ~ 2^{-3} N  Z_{\rm  [0]} (i/t) + ~ 2^{-3} N e^{\i \pi n } Z_{\rm [1]} (i/t) ~ ] \times
q^{4 \alpha^{\prime}\pi^2 n^2 /\beta^2} \cr
 =&& - \beta^{-1} \sum_{n \in {\rm Z} } \left [ \beta^{-1} (2\alpha^{\prime} \pi^3 n^2)^{1/2}
  + \beta (2\pi \alpha^{\prime})^{-1/2}  \right ] \cdot \rho_{\rm high} \quad ,
\label{eq:freeIpph}
\end{eqnarray}
where $\rho_{\rm high}$ is a constant independent of temperature.
Thus, the canonical ensemble of type I$^{\prime}$ strings is also
holographic, displaying the $T^2$ growth in free energy at high
temperatures characteristic of a {\em two-dimensional} field
theory!

\vskip 0.1in It is helpful to verify the corresponding scaling
relations for the first few thermodynamic potentials. The internal
energy of the free type I$^{\prime}$ string ensemble takes the
form:
\begin{eqnarray}
 U =&& - \left ( {{\partial W}\over{\partial \beta }} \right )_V
 \cr =&&
 \half L^9
(4\pi^2 \alpha^{\prime})^{-9/2} \int_{0}^{\infty} {{dt}\over{8t}}
e^{ -\beta^2t /2\pi \alpha^{\prime} }{{(2\pi t )^{-9/2}}\over{
  \eta(it)^{8}}} \cdot \sum_{n \in {\rm Z} } Z_{\rm open}
{{4 \pi t}\over{\beta}}  \left ( {{\beta^2}\over{4\pi^2
\alpha^{\prime}}} -
   {{ 4 \alpha^{\prime} \pi^2 n^2}\over{\beta^2}}  \right )
q^{\alpha^{\prime}\pi^2 n^2 /\beta^2 }
 , \cr &&
\label{eq:entermn}
\end{eqnarray}
where ${\rm Z}_{\rm open}$ is the factor in square brackets in the
expression in Eq.\ (\ref{eq:freeIp}). Unlike the heterotic string
ensemble, $U(\beta)$ no longer vanishes at the self-dual
temperature, $T_c$$=$$1/2 \pi\alpha^{\prime 1/2}$, since there are
no winding modes in the thermal spectrum.

\vskip 0.1in The analyticity of infinitely many thermodynamic
potentials in the vicinity of the critical point can be
demonstrated as for the heterotic string ensemble. We define:
\begin{equation}
[d t ] \equiv \half  L^9 (4\pi^2 \alpha^{\prime})^{-9/2} \left [
{{dt}\over{8t}} e^{-\beta^2 t /2\pi \alpha^{\prime}} {{(2\pi t
)^{-9/2}}\over{
  \eta(it)^{8}}} \cdot Z_{\rm open}
\right ] ,
\quad y(t;\beta)
   \equiv 2 \pi t \left ( {{\beta^2}\over{4\pi^2
\alpha^{\prime}}} +
   {{4\alpha^{\prime} \pi^2 n^2}\over{\beta^2}}
  \right ) \quad ,
\label{eq:varsn}
\end{equation}
and denote the $m$th partial derivative with
respect to $\beta$ at fixed volume by
$W_{(m)}$, $y_{(m)}$.
The higher derivatives of
the generating functional
now take the simple form:
\begin{eqnarray}
W_{(1)} =&& \sum_{n \in {\rm Z}} \int_{0}^{\infty} [dt] e^{-y}
(-y_{(1)})
%, \quad
\cr
W_{(2)} =&&
 \sum_{n \in {\rm Z} } \int_{0}^{\infty} [dt] e^{-y}
(-y_{(2)} + (-y_{(1)})^2 )
\cr
%, \quad
W_{(3)} =&& \sum_{n \in {\rm Z} } \int_{0}^{\infty} [dt] e^{-y}
(-y_{(3)} - y_{(1)} y_{(2)} +  (-y_{(1)})^3 ) \cr
 \cdots =&& \cdots
\cr W_{(m)} =&& \sum_{n \in {\rm Z} } \int_{0}^{\infty} [dt]
e^{-y} (-y_{(m)} - \cdots +  (-y_{(1)})^m ) \quad .
\label{eq:effmn}
%\end{equation}
\end{eqnarray}
Referring back to the definition of $y$, it is easy to see that the
generating functional and, consequently, the full set of thermodynamic
potentials is analytic in $\beta$. Notice that third and higher derivatives
of $y$ are determined by the momentum modes alone:
\begin{equation}
y_{(m)} = (-1)^m n^2 {{(m+1)! }\over{\beta^{m+2}}} , \quad m \ge 3 \quad .
\label{eq:dersn}
\end{equation}
Explicit expressions for the first few
thermodynamic potentials are as follows:
\begin{equation}
F = - {{1}\over{\beta}} W_{(0)} , \quad
U = - W_{(1)} , \quad
S = W_{(0)} - \beta W_{(1)} , \quad
C_V = \beta^2 W_{(2)} , \cdots \quad .
\label{eq:thermodln}
\end{equation}
The entropy is given by the expression:
\begin{equation}
S = \sum_{n \in {\rm Z} } \int_{0}^{\infty} [dt] e^{-y}
  \left [ 1 + 4\pi t (
{{ \beta^2}\over{4\pi^2 \alpha^{\prime}}} -
   {{4\alpha^{\prime} \pi^2 n^2}\over{\beta^2}} )
\right ] \quad ,
\label{eq:entropy1n}
\end{equation}
For the specific heat at constant volume, we have:
\begin{equation}
C_V =  \sum_{n \in {\rm Z}} \int_{0}^{\infty} [dt] e^{-y}
  \left [   16 \pi^2 t^2 (
{{\beta^2}\over{4\pi^2 \alpha^{\prime}}} -
   {{4\alpha^{\prime} \pi^2 n^2}\over{\beta^2}}
)^2
   - 4\pi t (
{{ \beta^2}\over{4\pi^2 \alpha^{\prime}}} +
   3 {{\alpha^{\prime} \pi^2 n^2}\over{\beta^2}} )
\right ] .
\label{eq:spcn}
\end{equation}
In summary, the free type I$^{\prime}$ string ensemble displays
holographic behavior at high temperatures characterized by a free
energy scaling as $\beta^{-2}$!

\subsection{An Order Parameter for a Gauge Theory Transition?}

\vskip 0.1in In this subsection, we will examine a plausible order
parameter for a phase transition in the gauge sector of the
theory. The calculation that follows relies on the well-known fact
that the sub-string-scale dynamics of string theory can be probed
by D0branes: pointlike topological string solitons whose mass
scales as $1/g$, in the presence of external background fields
\cite{dbrane,dkps,zeta}. In the free string limit the pointlike
D0branes behave like analogs of infinitely massive heavy quarks:
semiclassical, theoretical probes of the confinement regime. It is
well known that an order parameter signalling the thermal
deconfinement phase transition in a nonabelian gauge theory is the
expectation value of a closed timelike Wilson loop. We therefore
look for a signal of a thermal phase transition in the {\em short
distance behavior} of the pair correlator of timelike Wilson loops
in finite temperature string theory. We will use the fact that the
loop pair correlation function in string theory has a simple
worldsheet representation in terms of the path integral with {\em
fixed} boundary conditions \cite{cmnp,pair,zeta}. The short
distance behavior of this correlation function is dominated by the
shortest open strings, namely, the gauge sector of the massless
spectrum.

\vskip 0.1in Since we wish to focus on a property of the gauge
theory on the worldvolume of type I D9branes, we require {\em
both} fixed boundaries of the cylinder to lie on the same
orientifold plane, assumed to be a spatial distance $r$ apart in
the $X^9$ direction. Consider turning on an external constant
electric field, ${\cal F}_{09}$. The pair correlator of closed
timelike Wilson loops computes the Minkowskian time propagator of
a pair of electric sources lying in the worldvolume of D9branes
with external electric field, and with fixed spatial separation
$r$. In the T$_0$-dual type I$^{\prime}$ picture, the pointlike
sources are interpreted as being in relative motion in the
direction $X^0$ transverse to their spatial separation. As shown
in \cite{pair,zeta}, they experience an attractive binding
potential of the form $u^4 \alpha^{\prime 4}/r^9$ at zero
temperature, where the dimensionless parameter $u$ is defined as
$u$$=$${\rm tanh}^{-1} {\cal F}^{09}$. A systematic expansion in
field strength to all orders gives the small velocity, short
distance, corrections to the leading field-dependent binding
interaction between the sources \cite{pair,zeta}. The Wilson loops
represent the Euclidean worldlines of a pair of closely separated
semiclassical sources in slow motion.

\vskip 0.1in It is straightforward to extend this result to the
{\em finite temperature} correlation function and, for clarity, we
begin by focussing on the static limit, setting $u=0$. The path
integral expression for the pair correlator, ${\cal W}_2$, at
finite temperature takes the form \cite{cmnp,pair,zeta,decon}:
\begin{eqnarray}
{\cal W}_2(r,\beta)  =&& \lim_{r \to 0} \int_0^{\infty} dt {{e^{-
r^2 t/2\pi \alpha^{\prime} }}\over{\eta(it)^{8}}}
       \sum_{n\in {\rm Z}} q^{4\pi^2 n^2 \alpha^{\prime}/\beta^2 } \cr \quad && \quad \times [~
({{\Theta_{00}(it;0)}\over{\eta(it)}})^4
 -  ({{\Theta_{10}(it;0)}\over{\eta(it)}})^4  \cr
\quad &&\quad - e^{\i \pi n} \{
({{\Theta_{01}(it;0)}\over{\eta(it)}})^4 -
    ({{\Theta_{11}(it;0)}\over{\eta(it)}})^4 \} ~] .
\label{eq:pairc}
\end{eqnarray}
The static pair potential at short distances is extracted from the
dimensionless amplitude ${\cal W}_2$ as follows. We set ${\cal
W}_2$$=$$\lim_{\tau\to\infty} \int_{-\tau}^{+\tau} d\tau
V[r(\tau),\beta]$, inverting this relation to express $V[r,\beta]$
as an integral over the modular parameter $t$. Consider a $q$
expansion of the integrand, valid in the limit $r$$\le$$2\pi
\alpha^{\prime}$, $t$$\to$$\infty$, where the shortest open
strings dominate the modular integral. Retaining the leading terms
in the $q$ expansion and performing explicit term-by-term
integration over the worldsheet modulus, $t$, \cite{pair,zeta},
isolates the following short distance interaction \cite{decon}:
\begin{eqnarray}
V(r,\beta) =&&  (8\pi^2 \alpha^{\prime})^{-1/2} \int_0^{\infty} dt
e^{- r^2 t/2\pi \alpha^{\prime} } t^{1/2} \cr \quad &&\quad \times
\sum_{n\in {\rm Z}} ( 16 - 16 e^{i \pi n } ) q^{4\alpha^{\prime}
\pi^2 n^2 /\beta^2} + \cdots \cr \quad =&&2^4 {{1}\over{r(1+ {{
  r_{\rm min.}^2 }\over{r^2}} {{\beta_C^2 }\over{\beta^2 }} )^{1/2} }} + \cdots
\label{eq:static}
\end{eqnarray}
where we have dropped all but the contribution from the $n$$=$$1$
thermal mode in the last step. We have expressed the result in
terms of the characteristic minimum distance scale probed in the
absence of external fields, $r_{\rm min.}$$=$$2\pi \alpha^{\prime
1/2}$, and the closed string's self-dual temperature $T_C$.
Consider the crossover in the behavior of the interaction as a
function of temperature: at low temperatures, with $(r/r_{\rm
min.})\beta$$>>$$\beta_C$, we can expand in a power series. The
leading correction to the inverse power law is $O(1/\beta^2 r^3)$.
At high temperatures with $(r/r_{\rm min.}) \beta$$<<$$\beta_C$,
the potential instead approaches a constant independent of $r$. We
note the characteristic signal of the onset of the holographic
high temperature phase where the order parameter approaches a
constant {\em independent of $r$}. This is reminiscent of the
usual signal for confinement, but we should emphasize that we are
studying the short distance behavior of the Wilson loop
correlator. The short distance regime is not usually accessible in
standard gauge theory calculations; inferring the low energy field
theory result from the full string theory calculation has enabled
exploration of this phenomenon.

\vskip 0.1in Notice that at the crossover between the two phases
the potential is a precise inverse power law. Notice also that the
overall coefficient of the potential is independent of the
spacetime dimensionality of the gauge fields: the factor of $2^4$
is related to the critical dimension of the type I$^{\prime}$
string theory--- not the dimensionality of the worldvolume of the
Dpbrane in question.

\vskip 0.1in It is rather interesting to compare these two
observations with an old conjecture in the gauge theory
literature. Peskin has argued \cite{peskin} that when the
deconfining phase transition in a renormalizable gauge theory is
second order or is, more generally, a continuous phase transition
without discontinuity in the free energy, {\em the heavy quark
potential must necessarily take the scale-invariant form, $C/r$,
with $C$ a constant independent of spacetime dimension}. As
mentioned above, the short distance asymptotics of the Wilson loop
correlator is dominated by a result in {\em gauge theory} alone.
Moreover, perturbative string theory is fully renormalizable, and
its low energy limit is a {\em renormalizable} gauge theory.
Finally, the phase transition we have observed is a continuous
phase transition as follows from the analyticity properties
demonstrated earlier.

\vskip 0.1in The pair potential between semiclassical point
sources is indeed found to take the scale-invariant inverse power
form at criticality. And the coefficient of the potential is
independent of the dimensionality of the worldvolume of the
Dbrane, in other words, the spacetime dimension of the gauge
theory. Thus, our results are found to be in satisfying accord
with the intuition in Peskin's conjecture \cite{peskin}. But it
should be noted that the original conjecture was not made in the
context of thermal deconfinement. Nor was there a well-developed
notion of the short distance regime of the \lq\lq potential".
Hopefully, future work will bring further insight into these
issues.

\vskip 0.1in It is rather straightforward to include in this
result the dependence on an external electromagnetic field. From
our previous works \cite{pair,zeta,decon}, the result above is
modified by the simple replacement: $r_{\rm
min.}$$\to$$2\pi\alpha^{\prime 1/2}u$, where ${\cal
F}^{09}$$=$${\rm tanh}^{-1}u$ is the electric field strength. The
transition temperature in the presence of an external field,
$T_d$$=$$u T_C$, is consequently {\em lower} than the closed
string's self-dual temperature.

\section{Conclusions}

\vskip 0.1in As emphasized by us in \cite{bosonic}, while a
point-particle field theory with an exponential growth in the
density of states is indeed expected to exhibit a Hagedorn phase
transition, {\em there is no sign of such a transition in the
canonical ensemble of free strings}. Thus, by the usual procedure
of taking the $\alpha^{\prime}$$=$$0$ low energy field theory
limit, we can also infer {\em the absence of such a transition in
the zero coupling limit of the corresponding low energy field
theory}. In particular, we see no signal of a Hagedorn phase
transition at zero coupling in the nonabelian gauge theories at
finite temperature obtained in the $\alpha^{\prime}$$=$$0$ limit
of our string theory calculation: the free energy of the gauge
theory as inferred from the string theory result is both finite,
and free of divergences, at all temperatures. Is this property
unique to the anomaly-free gauge-gravity theories explored in
infrared-finite perturbative string theories, or does it hold more
generally for nonabelian gauge theory? It would be nice to test
this result from an independent investigation of the Hagedorn
transition in the zero coupling limit of gauge theories. Full
insight into the significance of the order parameter for the
thermal phase transition in the finite temperature gauge theory
demonstrated in section 4.3 also remains open for future work.

\vskip 0.1in Our results provide concrete evidence of the {\em
holographic} nature of perturbative string theory at high
temperatures \cite{aw,sussk,polchinskibook,bousso}: the growth of
the free energy at high temperatures is that of a {\em
two-dimensional} field theory. We have identified the duality
phase transition in the canonical ensemble of free strings in each
case--- heterotic $E_8$$\times$$E_8$ and ${\rm Spin(32)/Z_2}$,
type I and type I$^{\prime}$, or type IIA and IIB, as belonging to
the universality class of the Kosterlitz-Thouless transition
\cite{kt}, characterized by the analyticity in temperature of an
{\em infinite} hierarchy of thermodynamic potentials. Notice that
the precise manifestation of the thermal duality transition is
different in each case, and that it acts like a map between a pair
of supersymmetric string theories. Our results for the canonical
ensemble of type I strings are especially intriguing. As
demonstrated unambiguously in an accompanying paper using the
worldsheet renormalization group and g-theorem \cite{relevant,al},
we have shown that the nonsupersymmetric ground state described in
section 4 has both vanishing one-loop vacuum energy, and vanishing
dilaton tadpole. Supersymmetry is spontaneously broken by thermal
effects, except in the closed oriented sector of the theory.

\vskip 0.1in We would like to reiterate a point originally made by
us in \cite{decon}: since a thermal duality transformation is
nothing other than a Euclidean timelike T-duality transformation,
and the action of spacelike T-dualities among the six
ten-dimensional perturbative string theories is extremely
well-established \cite{polchinskibook}, it is {\em not} possible
to modify the precise role of thermal duality transformations as
used in this paper without also violating Lorentz invariance. We
offer this as a cautionary comment on recent conjectural
applications of thermal duality.

\vskip 0.05in An important general conclusion from this analysis
is that, from a physics standpoint, and in the absence of Dbrane
sources, the type II superstring theories and their corresponding
low-energy limits: pure supergravity theories with 32
supercharges, are a somewhat artificial truncation of the more
physical type II theories with Dbranes. These are string theories
with 16 or fewer supercharges, and they have both Yang-Mills
fields and gravity in common with the type I and heterotic string
theories. This insight has played a key role in my renewed focus
on theories with 16 supercharges in recent years. Recovering the
hidden symmetry algebra of supergravity theories with 32
supercharges as a special limit of the algebra of the theories
with 16 supercharges and Yang-Mills degrees of freedom, is a key
element of the proposal for M Theory given in \cite{mtheory}.

\vspace{0.2in} \noindent{\bf Acknowledgements:} I thank C. Bachas,
A. Dhar, J. Distler, M. Fukuma, P. Ginsparg, J. Harvey, G.
Horowitz, S.\ Kachru, H. Kawai, D. Kutasov, M. Peskin, J.
Polchinski, G. Shiu, B. Sundborg, H. Tye, and E. Witten for early
comments. This research was supported in part by the award of
grant NSF-PHY-9722394 by the National Science Foundation under the
auspices of the Career program. This paper was updated at the
Aspen Center for Physics, following the appearance of
\cite{relevant}. I am grateful to Hassan Firouzjahi for pointing
out a minor, but embarrassing, typo in the discussion of the
thermal tachyon spectrum of the pure type II closed string
ensemble in an earlier version of this work. I would like to thank
S. Abel, K. Dienes, H. Firouzjahi, M.\ Kleban, and E.\ Mottola for
their questions and interest, and Scott Thomas for the invitation
to present this work at the {\em Cosmic Acceleration} workshop.

\vskip 0.5in
\noindent{\bf Note Added (Sep 2005):} Many of the points made in this
paper are either extraneous, or incorrect in the details, although the broad
conclusions do stand. Namely, that there is no self-consistent type II 
superstring ensemble, in the absence of a Yang-Mills gauge sector.
The fact that both heterotic and type I theory have equilibrium canonical
ensembles free of thermal tachyons; a crucial role is played by the
temperature dependent Wilson line wrapping Euclidean time. In addition to
topics covered in hep-th/0105244, we have the discussion from hep-th/0208112
 of the type IB-I$^{\prime}$
 open and closed string
ensembles, including the
evidence for an order parameter for the unusual duality phase transition in
this theory \cite{decon}. A step
forward is the correction that the thermal tachyon spectrum in the type
II theories covers the full temperature range, down to T = 0. Note that the
pressure of the heterotic and type I string ensembles equals the negative
of the vacuum energy density, incorrectly stated in my previous papers; I
thank H.\ Firouzjahi for pointing this out.
The remaining errors have to do with thermal mode number dependent
phases that are not modular invariant, which I only became aware of in
hep-th/0506143.

\end{document}